\documentclass[sigconf]{acmart}
\usepackage{amsmath} 
\AtBeginDocument{%
  \providecommand\BibTeX{{%
    \normalfont B\kern-0.5em{\scshape i\kern-0.25em b}\kern-0.8em\TeX}}}

\setcopyright{acmcopyright}
\copyrightyear{2020}
\acmYear{2020}
\acmDOI{10.1145/1122445.1122456}

\acmConference[Woodstock '18]{Woodstock '18: ACM Symposium on Neural
  Gaze Detection}{June 03--05, 2018}{Woodstock, NY}
\acmBooktitle{Woodstock '18: ACM Symposium on Neural Gaze Detection,
  June 03--05, 2018, Woodstock, NY}
\acmPrice{15.00}
\acmISBN{978-1-4503-XXXX-X/18/06}

\begin{document}

\title{TROPPO LoRa: \\
TROPospheric Personal Observatory using LoRa signals}


\author{Marco Zennaro}
\email{mzennaro@ictp.it}
\orcid{1234-5678-9012}

\affiliation{%
  \institution{The Abdus Salam International Centre for Theoretical Physics}
  \city{Trieste}
  \state{Italy}
}

\author{Ermanno Pietrosemoli}
\email{ermanno@ictp.it}
\orcid{1234-5678-9012}
\affiliation{%
  \institution{The Abdus Salam International Centre for Theoretical Physics}
  \city{Trieste}
  \state{Italy}
}

\author{Marco Rainone}
\email{mrainone@ictp.it}
\affiliation{%
  \institution{The Abdus Salam International Centre for Theoretical Physics}
  \city{Trieste}
  \state{Italy}
}

\author{Daniele Trinchero}
\email{daniele.trinchero@polito.it}
\affiliation{%
  \institution{Politecnico di Torino}
  \city{Torino}
  \state{Italy}
}

\author{Mattia Poletti}
\email{mattia.poletti@polito.it}
\affiliation{%
  \institution{Politecnico di Torino}
  \city{Torino}
  \state{Italy}
}

\author{Giovanni Colucci}
\email{giovanni.colucci@polito.it}
\affiliation{%
  \institution{Politecnico di Torino}
  \city{Torino}
  \state{Italy}
}

\renewcommand{\shortauthors}{Zennaro and Pietrosemoli, et al.}

\begin{abstract}
  With the growth of LoRa deployments there are plenty  of anecdotal reports of very long wireless links, well beyond the line of sight. Most reports suggest that these links are related to anomalous tropospheric propagation. We developed a platform to study tropospheric links based on TheThingsNetwork, a popular LoRaWAN-based infrastructure. We present some preliminary results and call for the IoT community to participate in this radio propagation experiment.
\end{abstract}




\maketitle

\section{Introduction}
Electromagnetic waves in an homogenous medium travel in a straight line, so a sizable obstacle like the earth curvature will normally block the reception. There are several mechanisms that can change the speed and direction (the velocity) of a radio wave. Whenever there is a discontinuity in the transmission medium there will be a change in the direction of propagation. Depending on the specific type of discontinuity encountered the ray will undergo reflection, refraction, diffraction or scattering. A wave, characterized by many rays, could be affected by a combination of the above.

Reflection is an abrupt change of direction characterized by Snell's law; the angle of reflection equals the angle of incidence and the speed stays the same. The ray will not penetrate the second medium.
In refraction the ray changes direction and speed after penetrating the second medium, and the amount of direction change is proportional to the differences between the refraction indices of the two media. The refraction index \textit{n} of a medium  is the ratio between the speed of the electromagnetic wave in vacuum and the speed in the medium. 

\section{Tropospheric propagation}
The atmosphere between ground level and some 12 kilometers is called the troposphere, and in it  the pressure, temperature and humidity  decrease monotonically with altitude under normal conditions.

\begin{figure}[ht]
  \centering
  \includegraphics[width=\linewidth]{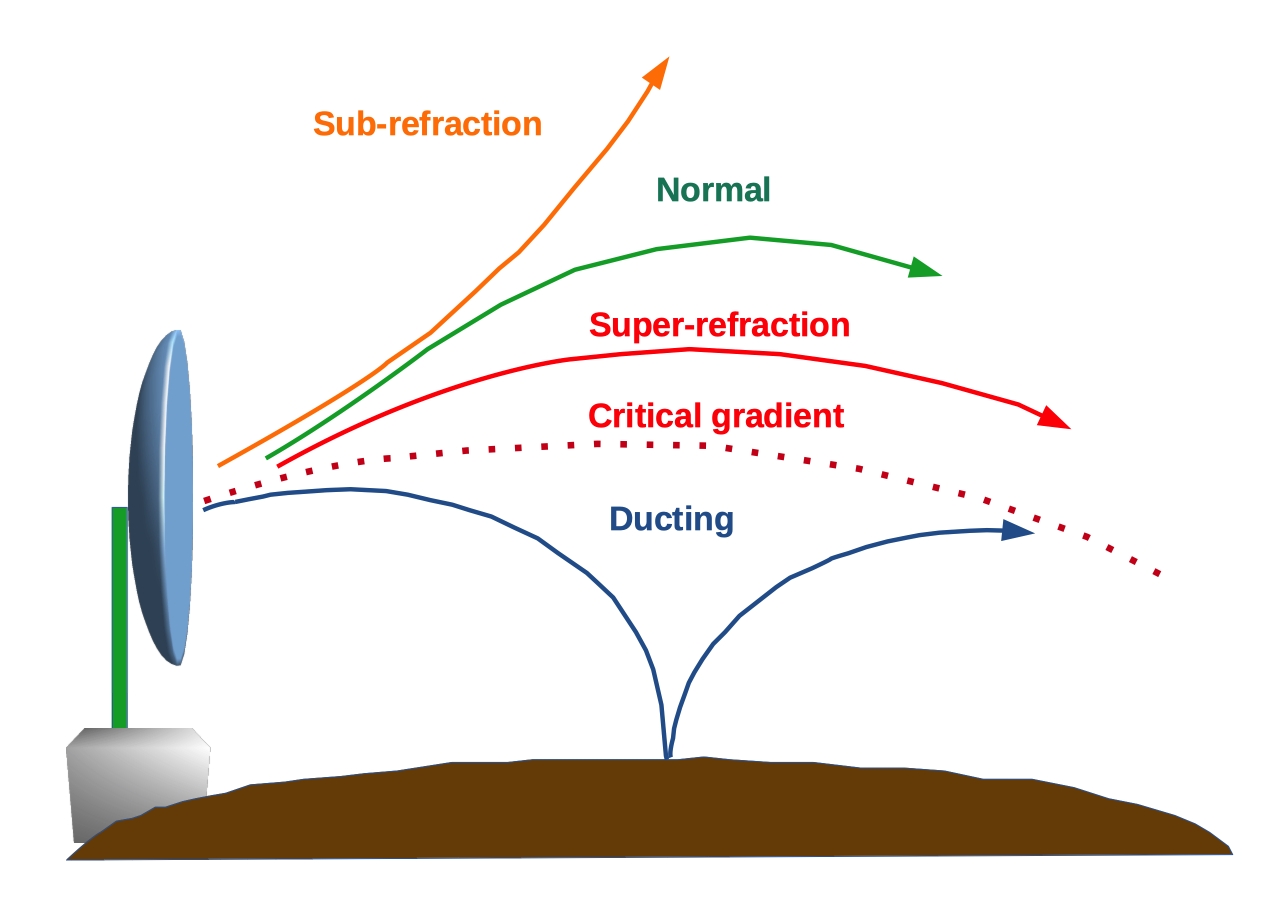}
  \caption{Disruptions on the temperature and humidity profiles in the troposphere cause anomalous propagation of different kinds.}
  \Description{Tropospheric ducting}
  \label{ducting}
\end{figure}

Disruptions on the temperature and humidity profiles in the troposphere on frequencies from about 50 MHz to 10000 MHz cause a  change in the refraction index that can significantly change the  propagation range as shown in Figure  \ref{ducting}.

Unusual variations of the humidity  and temperature of the air will create  a change of the refractive index that can result in a downward bending of the ray that will overcome the earth curvature in what is known as \textbf{super-refraction}\cite{series2019}. If  the refractivity gradient is very high, we can have a very long transmission range in what is called a \textbf{tropospheric duct}.

\begin{figure}[ht]
  \centering
  \includegraphics[width=\linewidth]{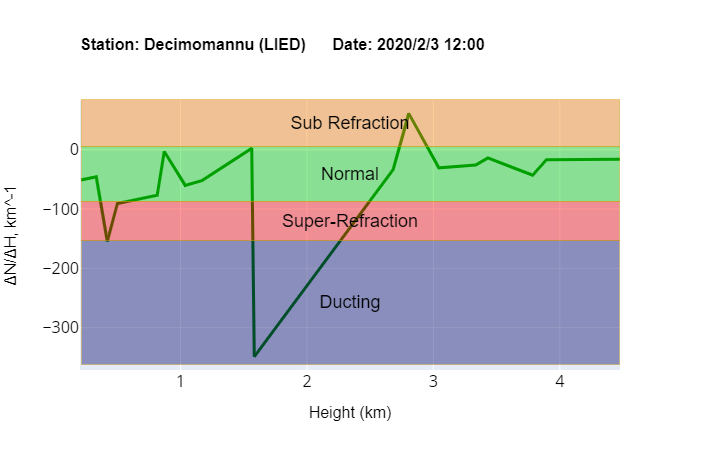}
  \caption{Refractivity gradient versus height. Four propagation conditions  are clearly distinguishable.}
  \Description{Refractivity gradient}
  \label{Refractivity}
\end{figure}


Since the refraction index of air is very close to 1, a new \textbf{refractivity} parameter \textit{N} is defined by N=(n-1)*$\ 10^{6}$\, allowing for the use of more manageable numbers. The value of refractivity N  at different heights can be calculated using the formula \eqref{eq:1} from  \cite{barclay2003propagation}, where P is atmospheric pressure in hPa, T is the temperature in kelvin and r is the relative humidity  in g/kg.  

\[N=77.6 *\frac{P}{T} + 3.73* 10^{5} \frac{r*P}{T^{2}*(622+r)} \tag{1} \label{eq:1} \]

The easiest way to obtain P, e and r at different elevations is from the meteorological radiosondes that are routinely launched world-wide \cite{durre20116b}.

As tropospheric propagation is caused by the \textbf{deviation}  from the normal profile of refractivity with altitude \cite{barclay2003propagation}, we use the  \textbf{gradient  $\Delta$N/$\Delta$h}, expressed in km$\ ^{-1}$\, to study such phenomena. 

In normal  conditions  0 < $\Delta$N/$\Delta$h  < -79 , the ray  has  as slightly greater  radius of curvature  than the Earth so it will reach a little beyond the optical horizon, as depicted in green in Figure 1. If  -79 < $\Delta$N/$\Delta$h < -157, the ray will have  a significantly greater radius, thus reaching considerable distances (red in Figure 1) before touching ground.

When  $\Delta$N/$\Delta$h = -157, the ray has the same radius as the earth and will stay at the same distance above the ideal ground (red dotted line in Figure 1). 
If  $\Delta$N/$\Delta$h  <  -157, the radius of the ray (in blue in Figure 1) is much smaller than that of the earth (in brown in Figure 1) and  will soon reach it reflecting upwards, then reflect again on the atmospheric disturbed layer. This pattern is repeated a number of times reaching distances of thousands of kilometers in what is known as a  \textit{tropospheric duct}. This is not very frequent and happens more often in trajectories over water which has excellent reflectivity.
If  $\Delta$N/$\Delta$h is positive, the ray  will bend upwards (orange in Figure 1) and  the radio horizon is shorter than the optical horizon.

Figure  \ref{Refractivity} shows $\Delta$N/$\Delta$h versus \textit{h}  from which the type of tropospheric propagation can easily be inferred.

Standard practice is to substitute the curved  rays trajectories by straight lines, drawn over an earth with a modified radius multiplied by what is known as the K factor \cite{platform}, were K= 1/[1+a*$\Delta$N/$\Delta$h] and \textit{a} is the true earth radius, 6371 km. In a normal atmosphere, K = 4/3 and the radio horizon is 1.33 greater than the optical horizon.

Reception  when the line of sight is blocked can also happen   when there is a sharp physical obstacle in the path  that  can \textbf{diffract} the radio waves in many directions, but the amount of energy in a particular direction will be heavily attenuated.
The attenuation in dB introduced by a sharp object can be calculated \cite{series2018propagation} using:

\[J(v)= 6.9 +20*log[\sqrt{(v-1)^2+1} +v - 0.1] \tag{2} \label{eq:2} \]    


where \textit{$\nu$} is a geometric parameter that depends on the wavelength,  the height and the angles of view and distances from each end-point to the obstacle. Diffraction from round shaped obstacles is much less pronounced.

Even in a normal troposphere, dust, rain, snow or local inhomogeneities can change the trajectory of a radio signal so that it can reach well beyond the earth curvature in what is known as \textbf{tropospheric scatter}. Scattering occurs when the wave encounters objects whose dimensions are of the order of magnitude of the propagating wavelength. The energy will be radiated in many different directions as opposed to what happens in specular reflections.



\section{LoRaWAN and The Things Network (TTN)}

The LoRaWAN architecture \cite{sornin2015lorawan} is composed of three elements: the end-nodes, the gateways and the servers. An end-node broadcasts data frames using the LoRaWAN protocol and any of the gateways is able to receive the packets. Gateways will blindly relay the data to a Network Server using any kind of IP connectivity. The Network Server examines a specific field in the frame to determine which application should a particular frame be sent to. Applications are hosted in an Application Server. 
 

 
 The wireless communication takes place through a single-hop link between the end-device and one or many gateways. In Europe \cite{devices2012radio} LoRa uses eight channels in the license-free radio frequency bands  from 863 to 870 MHz. Maximum  Equivalent Radiated Power (ERP) is 25mW (14 dBm) and the end-device transmit duty-cycle should be 0.1\% or 1.0\% per hour depending on the channel.  
 
 LoRa uses what is known as a "chirp" modulation \cite{chirp}, a technology developed for sonar and radar applications. It uses fixed amplitude linearly varying frequency signals that span the entire channel spectrum. The spreading factor (SF) represents the number of chips used to encode each bit of information independently on the adopted error correction scheme. LoRa in Europe operates with spreading factors from 7 to 12. SF7 is the shortest time on air, SF12 will be the longest. Each step up in spreading factor doubles the time on air to transmit the same amount of data. Longer time on air  results in less data transmitted per unit of time. Each step up in SF correlates to about 2.5 dB (theoretically, 3 dB) so SF12 will provide the highest sensitivity corresponding to the capability of detecting a signal up to 20 dB lower than the noise plus interference level, provided that the signal is above the receiver's sensitivity threshold.

\begin{figure}[ht]
  \centering
  \includegraphics[width=0.75\linewidth]{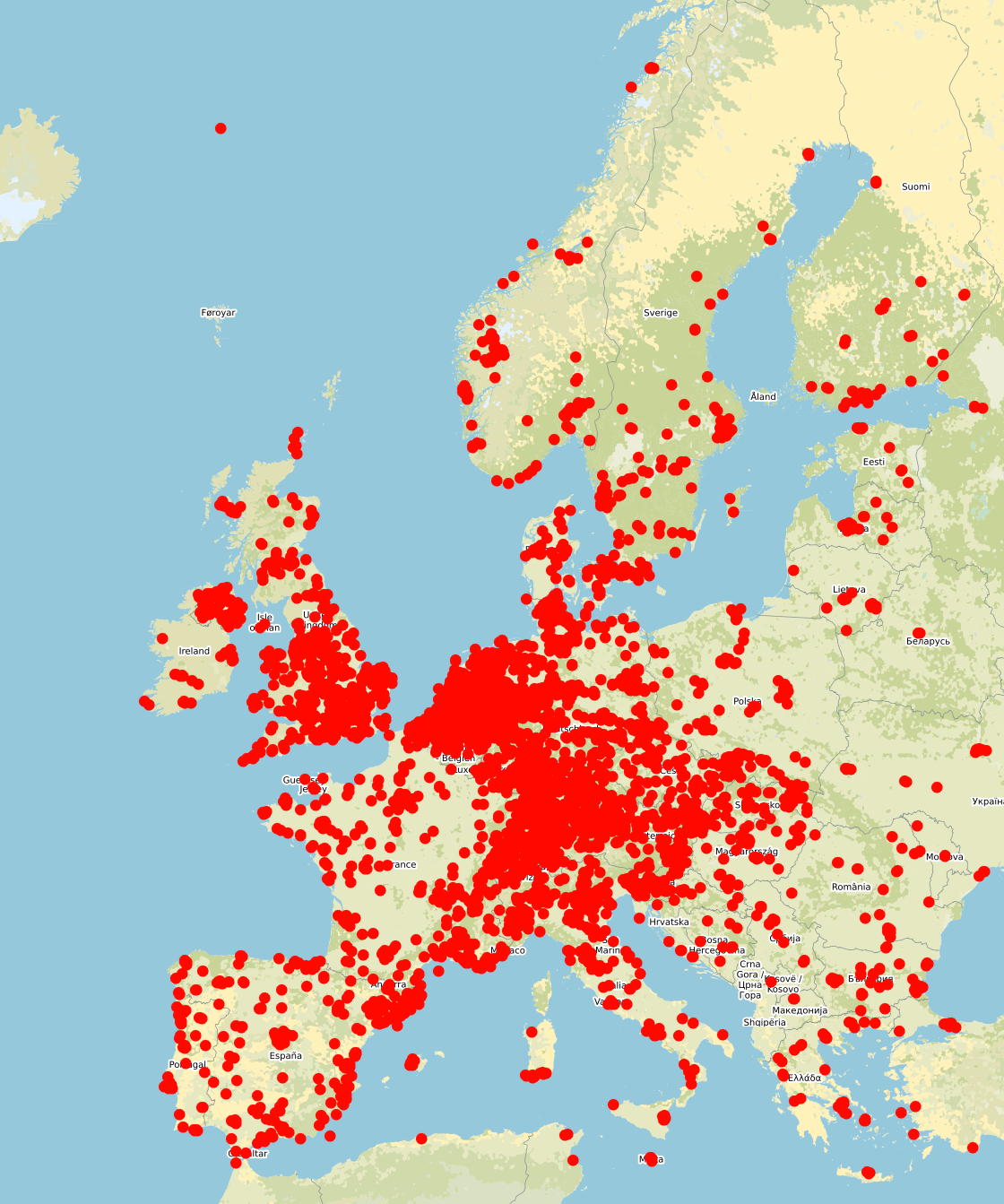}
  \caption{Map of TTN gateways in Europe.}
  \Description{Map of TTN gateways in Europe}
  \label{ttnitaly}
\end{figure}

 The Things Network (TTN) is an initiative \cite{giezeman2016things} to build a world-wide open source infrastructure to facilitate a public Internet of Things (IoT). TTN uses LoRaWAN to  build  IoT  networks.  LoRaWAN  provides  low-bandwidth communications  over  long  distances.  In  typical  applications IoT  nodes  send  few  bytes  of  data  every  few  minutes/hours over distances in the range of 1-10 kilometers. These technical characteristics allow the use of cheap devices that can operate for several  years  on  a  single  battery.  TTN  operates  by  allowing users  to  share  the  access  to  gateways,  which  play  the  role of  access  points  for  LoRaWAN  networks.  As  the  amount  of traffic generated by nodes is very limited, users are willing to share the access to their gateways and this allows the network to have a greater coverage. The Things Network, being a free public community LoRaWAN network, applies the fair use policy of  restricting uplink airtime to 30 seconds per day per node. There  are  now  over 11,000 TTN gateways  worldwide,  in more  than  100  cities. Figure \ref{ttnitaly} shows the distribution of gateways in Europe as an example.
 TTN manages both the Network and the Application Servers for the end nodes that are accordingly configured.
 
%
%
%
%
%

\section{The TROPPO platform}

The main  goal of the TROPPO platform \cite{platform} is to examine radio trajectories from end-nodes to TTN Gateways to understand if  anomalous propagation has ocurred. 

We developed probes (TP: TropoProbes) which periodically send data that can be received by any gateway in range. The same packet of 4 bytes is sent using different SFs, ranging from 7 to 12. Since we deployed the TPs we know their exact position and height above ground. These devices have been added to a TTN application. The physical devices are LoPy4 from Pycom \cite{lopy}, with an omnidirectional antenna. The output power has been set to 14 dBm since we estimate that the cable and connector losses compensate for the antenna gain so the total equivalent radiated power (ERP) is 14 dBm.

As we want to monitor more devices than just the ones we deployed, the platform is developed so that it can digest data coming from other end nodes as well. For that, we need to access only the metadata of the packets being sent in order to be able to add these nodes to the TROPPO platform.

The packets are also received by TTN gateways that are deployed by third parties. As the coordinates of the TTN gateways must be provided when registering it in TTN, we can determine the trajectory from a specific end-node to a given gateway.   

\begin{figure}[ht]
  \centering
  \includegraphics[width=\linewidth]{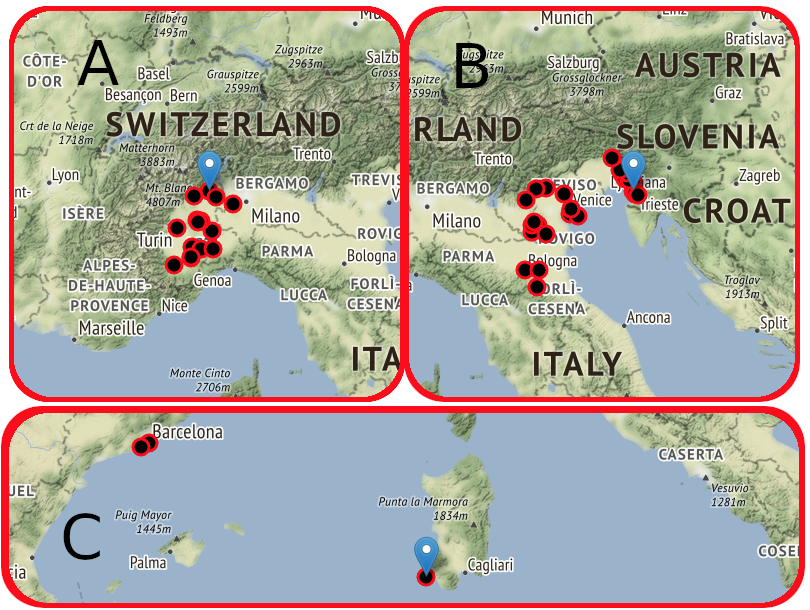}
  \caption{The blue balloons are locations of Tropo probes, the red circles are TTN gateways that have received packets from  probes. A: Gattinara probe,Piedmont, B ICTP probe, Trieste, C Carloforte probe, Sardinia. Red circles are TTN gateways that received their packets.}
  \Description{The blue balloons are locations of a Tropo Probe, the red circles are TTN gateways that have received packets from that Probe. A: Gattinara probe, B ICTP probe, Trieste, C Carloforte probe, Sardinia. Red circles are TTN gateways that received their packets.}
  \label{3sperimenti}
\end{figure}

The gateways forward all received frames to the TTN application server attaching a sequence number, timestamp and the value of the received signal strength indicator (RSSI). The information about the trajectory and the path attenuation can be easily deducted. In Figure \ref{3sperimenti} blue balloons are locations of a TPs, while red circles are TTN gateways that have received packets from them.

Our web platform allows the visualization of trajectories, shows the RSSI  and SNR received over time,  the SF distribution of received packets and depicts the terrain profile between the TP and the TTN gateway. Visualization can be done for data received in the last 1, 10 and 30 days and for all received data.

\section{Preliminary results}

\subsection{Trieste Tropo Probe}

The first TP, shown in  Figure \ref{3sperimenti}B, has been placed on the roof of the ICTP's Marconi lab building in Trieste, Italy, at 72 meters ASL. The packets have reached about 20 TTN gateways, including three gateways owned by us that worked as baseline for packet reception.

The farthest gateway we could reach is near Cesena, in Central Italy, at a distance of 235 km. This is indeed a case of anomalous propagation since the earth's curvature is completely blocking the line of Sight (LOS) at this distance.


The gateway received a total of 18521 packets in a three months time frame (from December 2019 to March 2020). Figure  5 shows the time distribution of packets received by the Cesena gateway. Some days no packets were received since anomalous propagation  depends on atmospheric conditions.

\begin{figure}[ht]
  \centering
  \includegraphics[width=\linewidth]{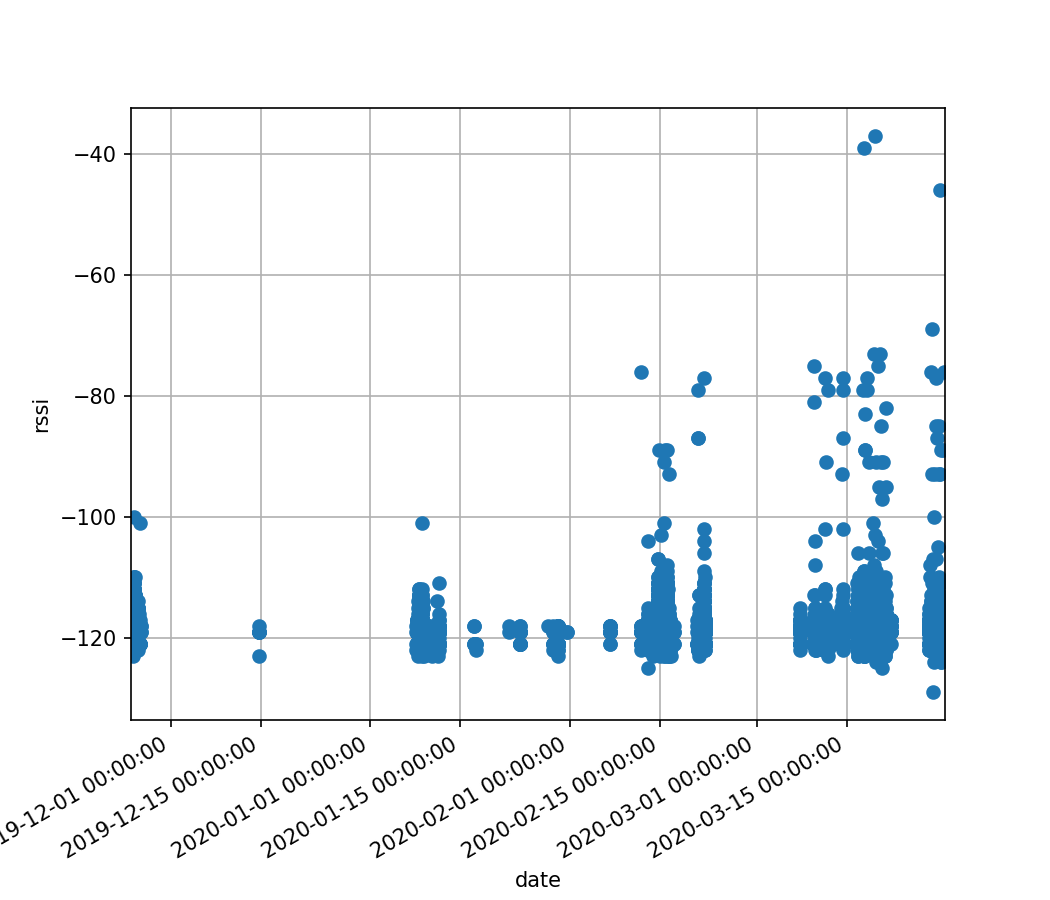}
  \caption{Distribution of packets received by  gateway in Cesena. Some days no packets were received since anomalous propagation  depends on atmospheric conditions.}
  \Description{Time distribution of packets received by the gateway in Cesena.}
  \label{timeTP1}
\end{figure}

\begin{figure}[ht]
  \centering
  \includegraphics[width=0.75\linewidth]{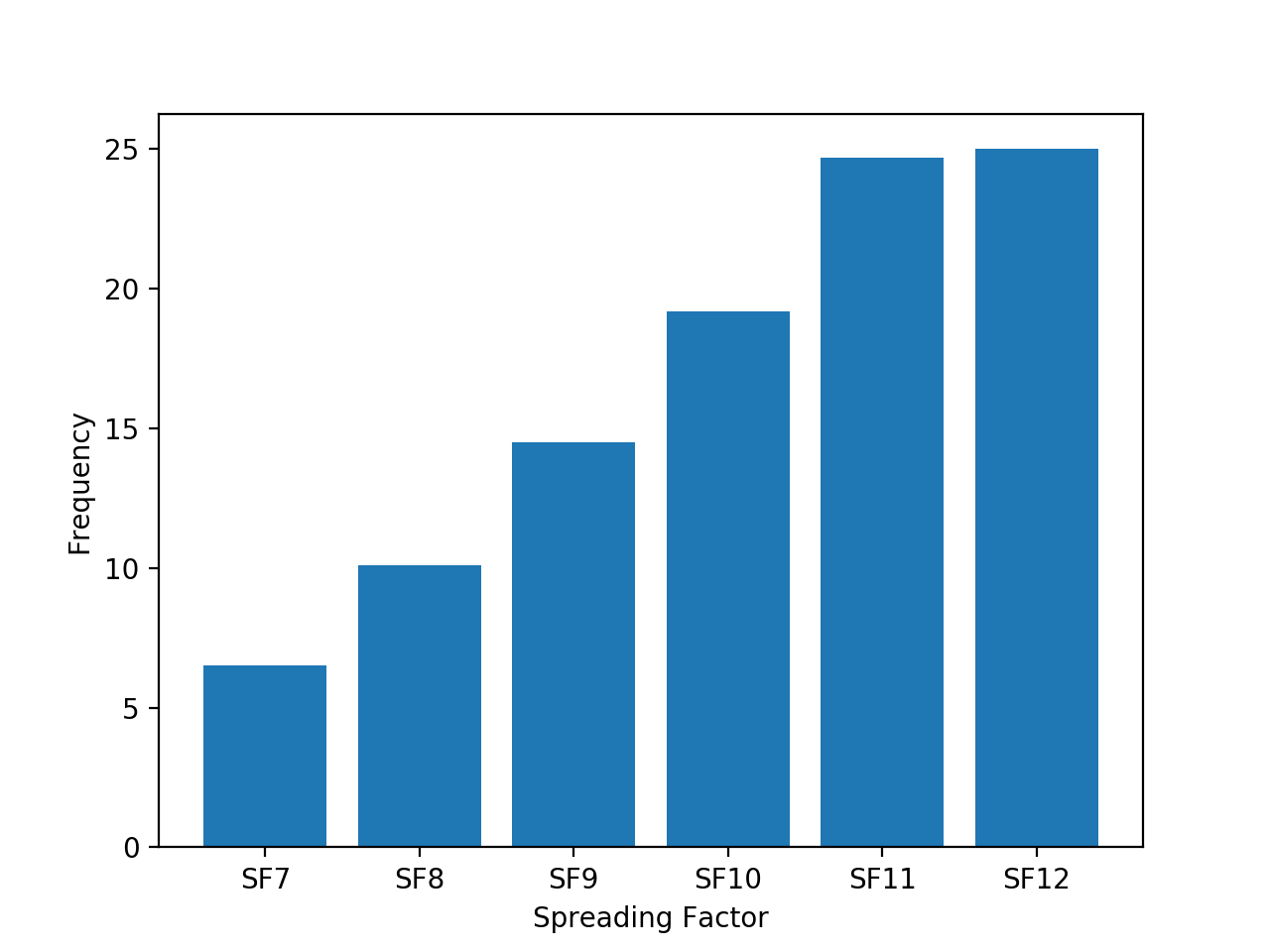}
  \caption{The distribution of SF shows that packets sent with SF 12, 11 and 10 have a higher chance of being received than the ones with lower SF.}
  \Description{The distribution of SF shows that packets sent with SF 12, 11 and 10 have a higher chance of being received than the ones with lower SF.}
  \label{SFTP1}
\end{figure}

The distribution of SF in Figure \ref{SFTP1} shows that packets sent with SF 12, 11 and 10 have a higher chance of being received than the ones with lower SF, since the higher SFs allow for reception at lower SNR. 

\subsection{Sardinia Anomalous Propagation}

A LoRaWAN device has been placed in the "U Tabarka" vineyards in the island of San Pietro, south Sardinia, by the iXem Labs \cite{colucci2017internet} in the framework of a smart agriculture project.  It is located at 50 meters ASL and at only 2 meters above ground. As this device has been designed for another application, it uses a fixed SF of 12. We are recycling the data sent for the purpose of analyzing the anomalous propagation.

The packets have reached three TTN gateways, one in Sardinia (the intended recipient, part of the same project) and two in Barcelona, Catalonia, at 573 and 584 km distance. Figure \ref{3sperimenti}:C shows as a blue balloon the location of the probe. The two red circles are TTN gateways which received packets beyond the optical horizon. 

One of the gateways in Barcelona has received 104 packets in a 1 month period (beginning of February 2020 to March 2020), while the second one has received only 3 packets in the same period. 



\begin{figure}[ht]
  \centering
  \includegraphics[width=\linewidth]{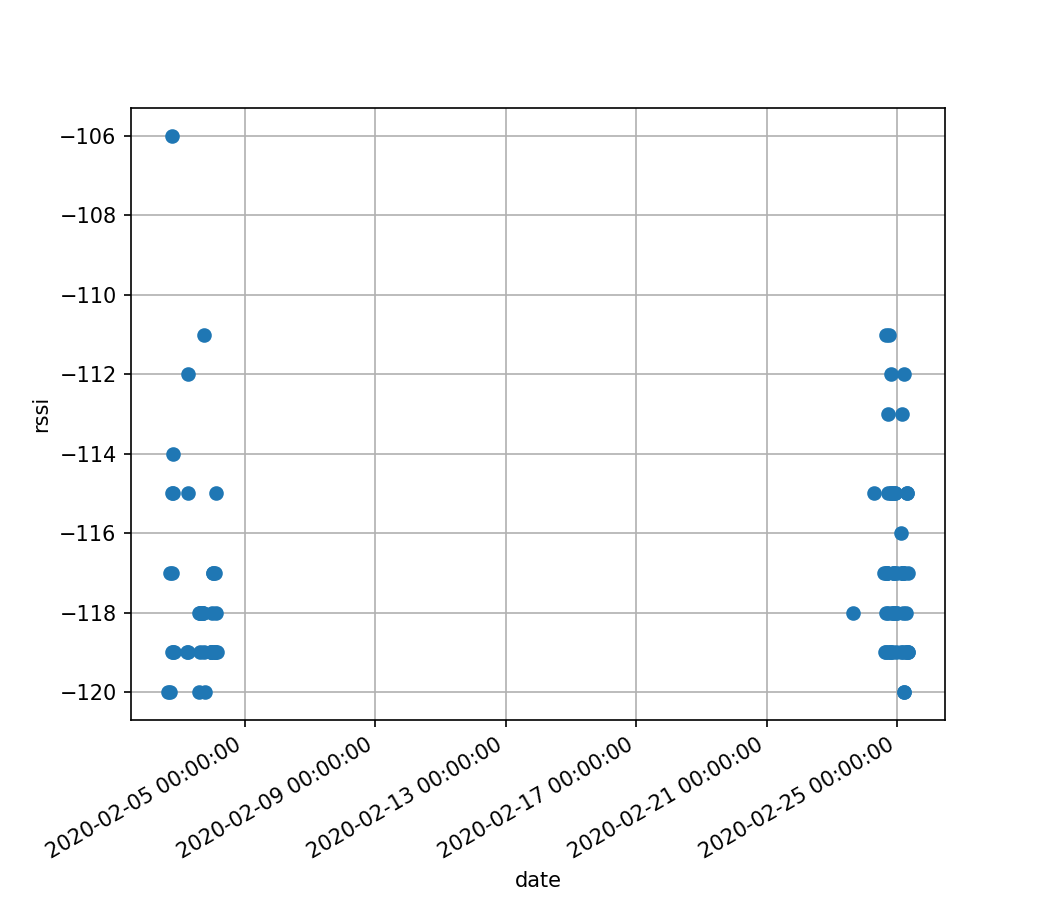}
  \caption{Packet received in Barcelona. The absence of packets from February 4, 2020 to  25, is due to the weather dependence of the tropospheric duct.}
  \Description{Notice the clustering of packets on February 3 and 25, 2020 and their absence in the intervening days, confirming the weather dependence of atropospheric duct.}
  \label{timeTP2}
\end{figure}

Distribution of packets received by the gateway in Barcelona is depicted in Figure \ref{timeTP2}. Notice the clustering of packets on February 3 and 25, 2020 and their absence in the intervening days, confirming the weather dependence of anomalous propagation.

\section{Discriminating the type of tropospheric propagation}

When very long LoRa trajectories are mentioned, it is not always clear which kind of anomalous propagation is involved. While anecdotes about very long links are abundant \cite{longlink}, it is important to use tools such as the ones we developed to be able to correctly adjudicate the specific type of propagation. Let's analyse the three cases described below.

\subsection{Super-refraction}

\begin{figure}[ht]
  \centering
  \includegraphics[width=\linewidth]{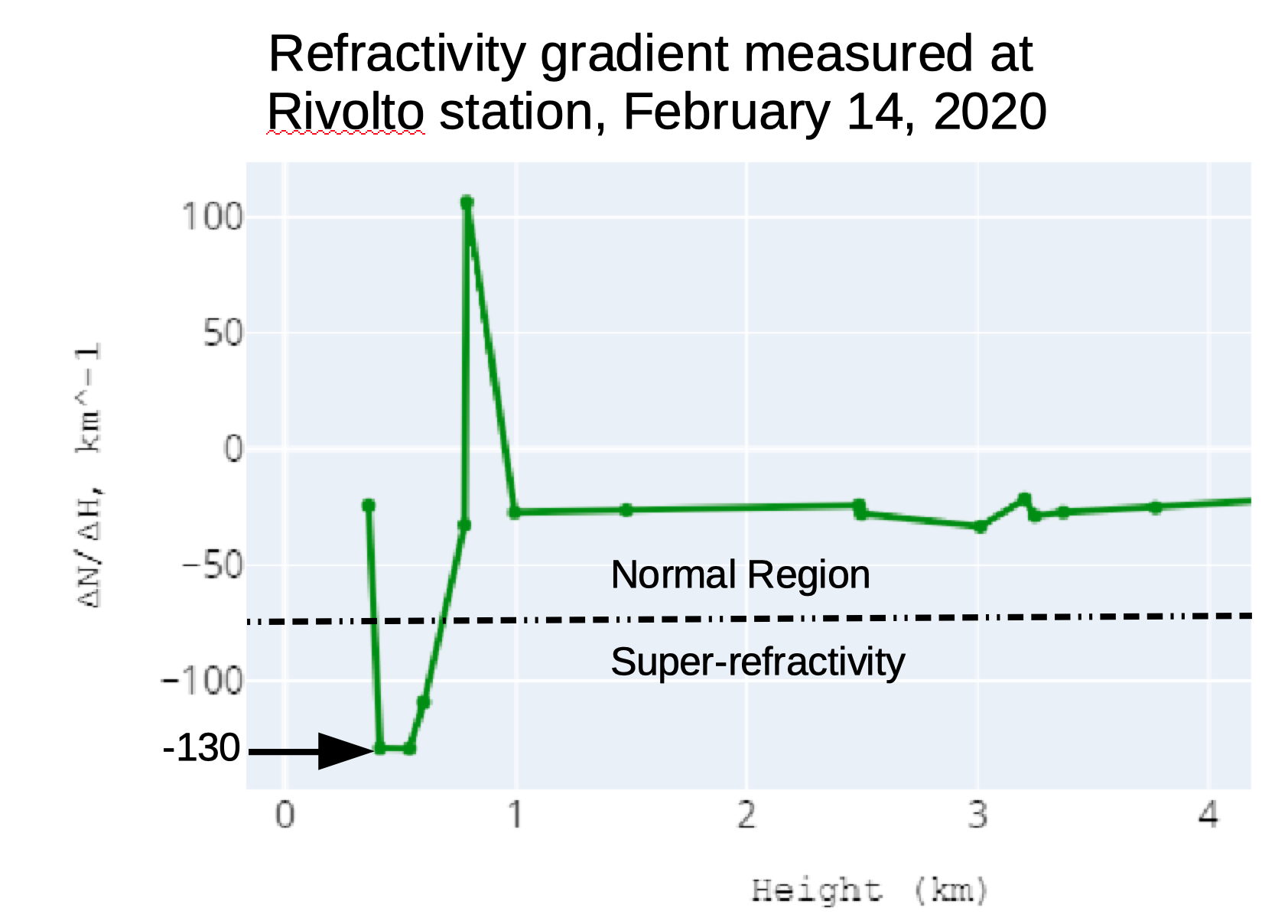}
  \caption{Refractivity gradient versus height at Rivolto Station. Values below -79 $ km  ^{-1}$ confirm super-refraction in the link between ICTP and Cesena.}
  \Description{ Rivolto Refractivity gradient}
  \label{Rivolto}
\end{figure}

A radiosonde was launched on February 14 from the meteorological station in Rivolto, 63 km from the ICTP building. With its data we calculated the refractivity gradient, shown in Figure \ref{Rivolto}, evidencing that it reached -130 $ km  ^{-1}$ at  0.6 km, thus confirming the presence of \textbf{super-refraction} a day in which we received packets in Cesena, as shown in Figure \ref{timeTP1}. Therefore a factor K = 5.85 must be applied to the radius of the earth in this case. 

Figure 9, made  with BotRf \cite{zennaro2016radio},  a telegram bot we built for planning RF links, confirms clearance over the terrain.

\begin{figure}[ht]
 \centering
  \includegraphics[width=\linewidth]{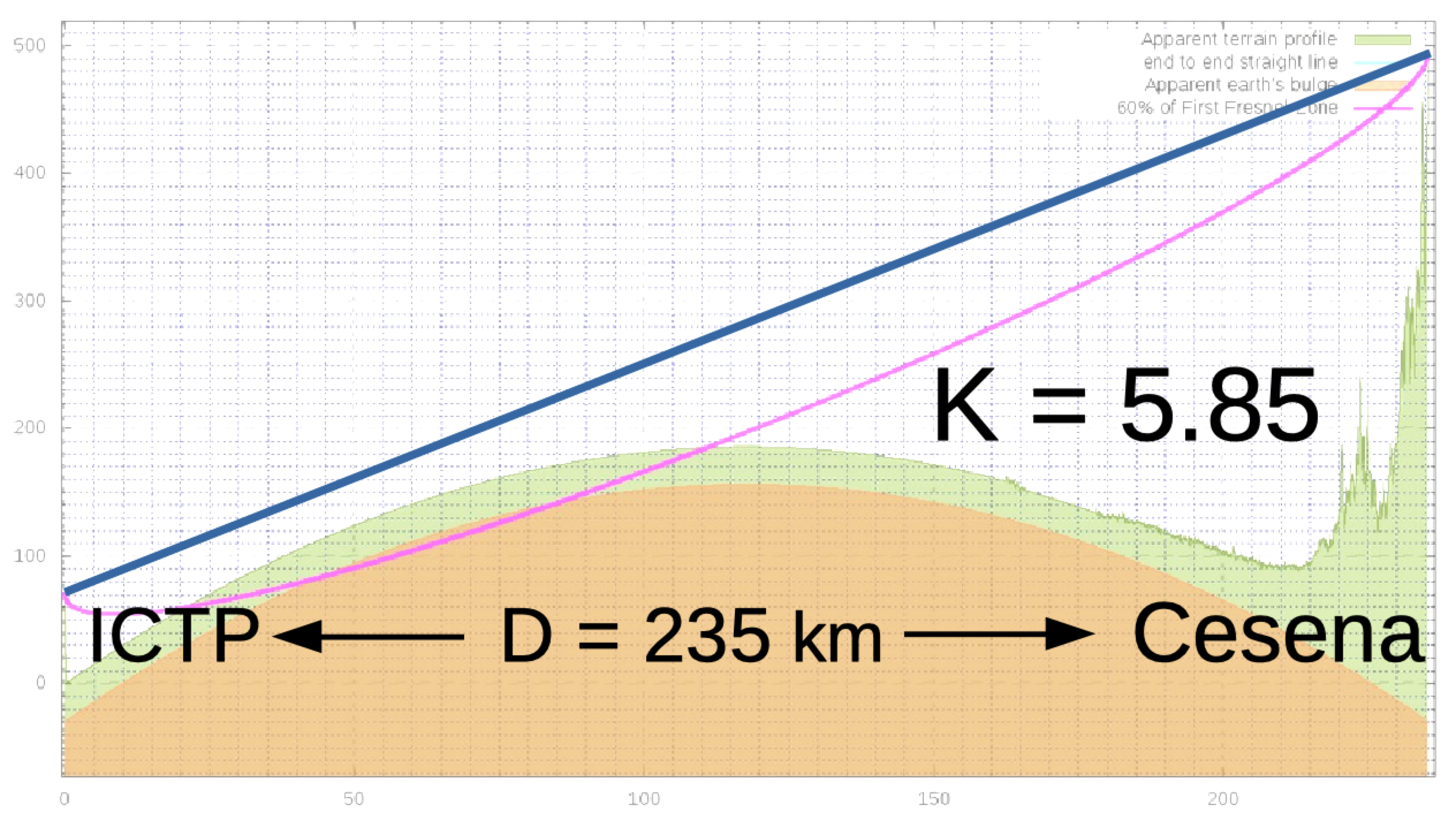}
  \caption{Terrain profile betweeen ICTP and Cesena drawn over an earth with the modified radius of curvature corresponding  to the refractivity gradient of -130 $ km  ^{-1}$, confirming the existence of a clear link.}

  \Description{Profile betweeen ICTP and Cesena.}
  \label{terrainICTP}
\end{figure}

\subsection{Tropospheric duct}

If we consider the long link from Sardinia to Catalonia, on February 3, 2020, 20 packets with SF 12 on a 125 kHz bandwidth were received at a Barcelona TTN gateway with RSSI spanning from -112 to -120 dB and SNR form -9.5 dB to -20.5 dB.

The same day a radiosonde was launched at Decimomannu Meteorological station, (50 km from Island of San Pietro),  which recorded atmospheric pressure, temperature and relative humidity at different heights, data used to draw Figure 2. Notice that $\Delta$N/$\Delta$h reaches -350 km$\ ^{-1}$\ confirming that we are in presence of a \textbf{tropospheric duct}.

\subsection{Diffraction}

\begin{figure}[ht]
  \centering
  \includegraphics[width=\linewidth]{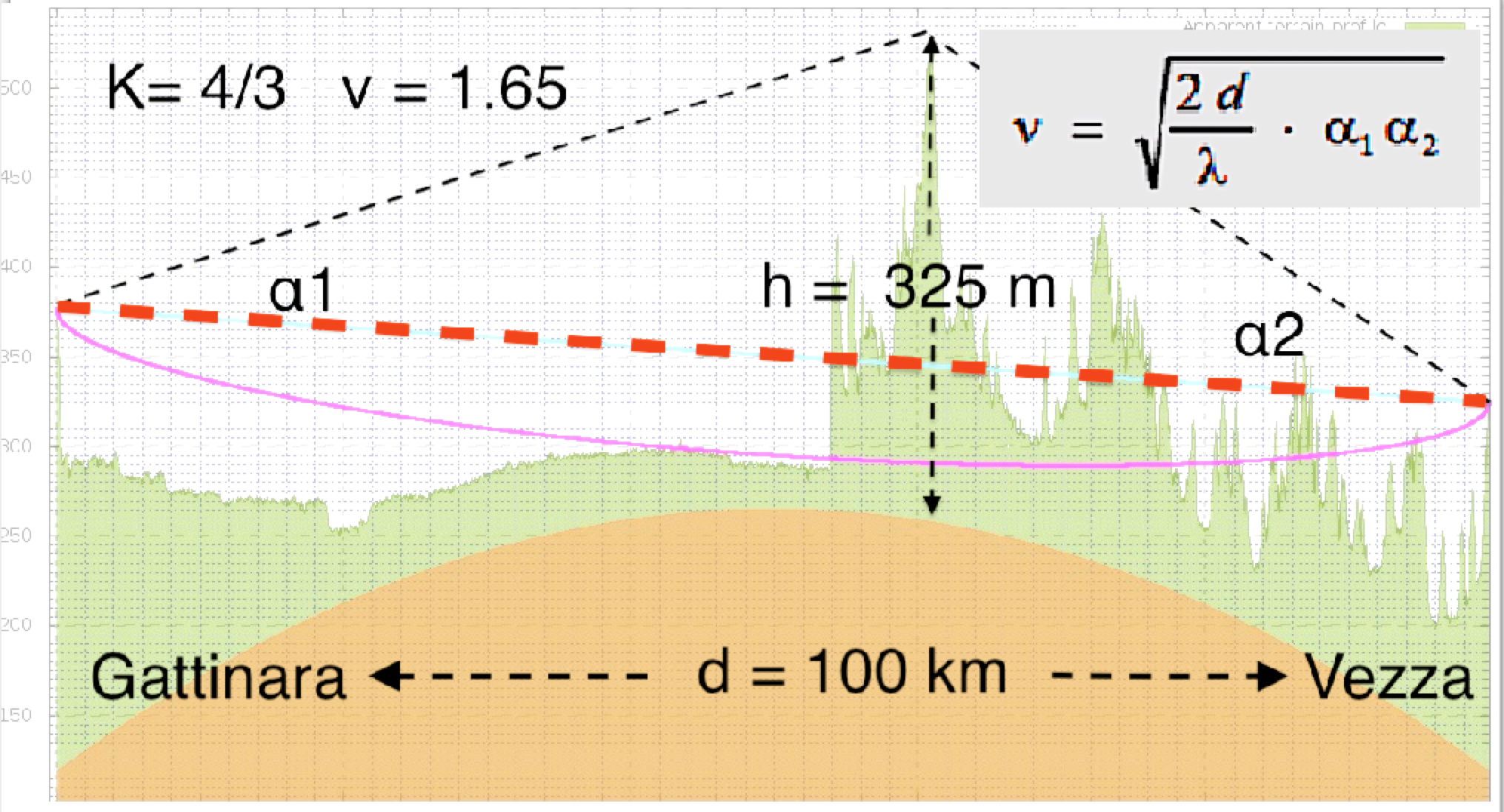}
  \caption{Profile Gattinara Vezza at k = 4/3. The sharp edge diffraction introduces an additional loss of 19 dB, but reception is still possible.}
  \Description{Profile Vezza}
  \label{ProfileVezza}
\end{figure}

Consider the TP shown in part A of Figure \ref{3sperimenti}. The blue balloon is a sensor installed by iXem Labs in the framework of a smart agriculture project at Gattinara in the Piedmont region of Italy. The LoRaWAN signal reaches about 15 TTN gateways. The gateway at Vezza d'Alba  received 5 packets on February 10, 2020, when the radiosonde data from the station at Cuneo (35 km away) evidenced normal refractivity as shown in the green trace of Figure 12. Yet the profile is blocked, so this is a case of diffraction  by a sharp ridge,  where the level of the received signal is calculated by adding to the free space loss an additional term to account for the diffraction loss \cite{series2018propagation}, calculable using equation (2).

The  parameter  $\nu$  is obtained from the formula on Figure  \ref{ProfileVezza}.
Applying this formula to the geometry of this link in normal refractivity conditions, (k = 4/3) results in  $\nu$ = 1.65. Entering this value in equation 2 results in a diffraction loss of 17 dB. Since the FSL at 868 MHz over 100 km is 131 dB, the total attenuation is 148 dB, so the calculated received signal is 14 -148 = -134 dBm, easiliy detectable.

\begin{figure}[ht]
  \centering
  \includegraphics[width=\linewidth]{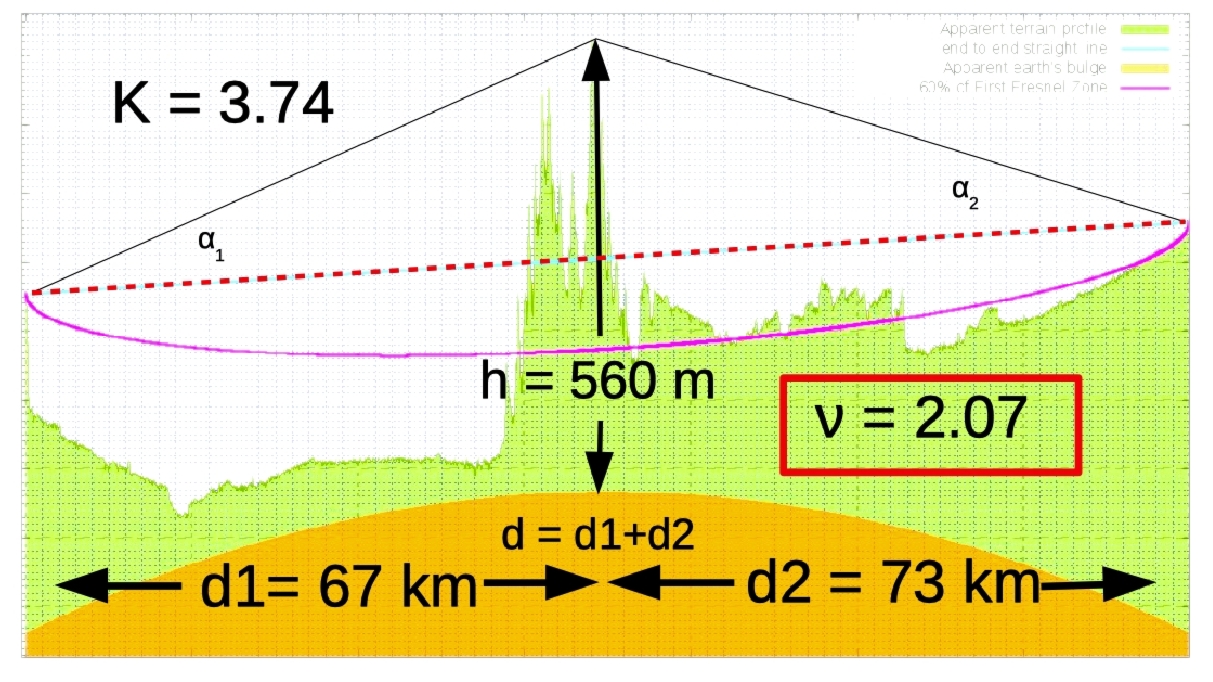}
  \caption{There is no line of sight between the probe in Gattinara and the TTN gateway in Cuneo. Diffraction  combined with  super-refraction is the mechanism involved.}
  \Description{There is no line of sight between the probe in Gattinara and the TTN gateway in Cuneo. Diffraction  combined with  super-refraction is the mechanism involved.}
  \label{gattinara}
\end{figure}

\begin{figure}[ht]
  \centering
  \includegraphics[width=\linewidth]{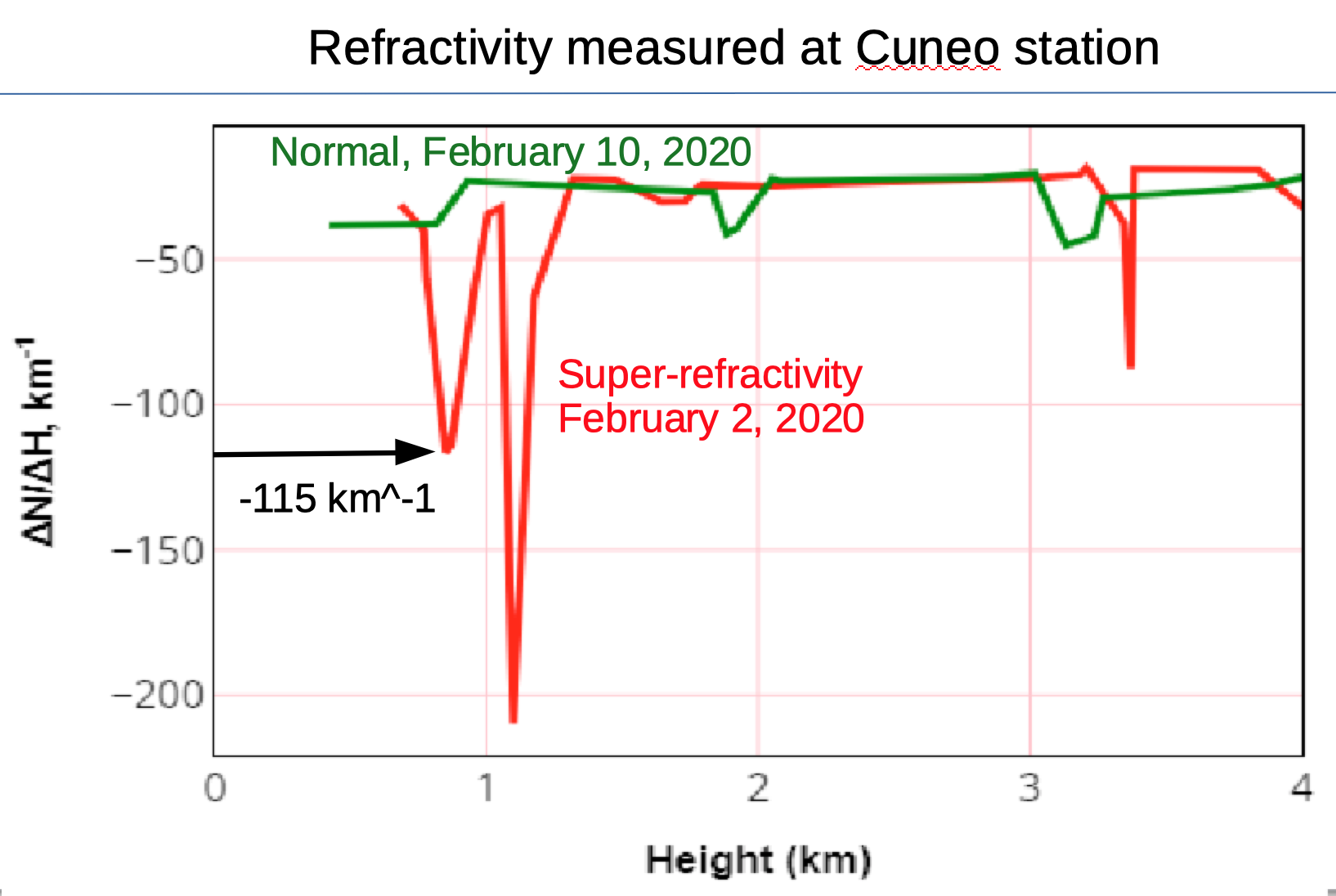}
  \caption{Refractivity gradient versus height at Cuneo station. In green a day with no reception, where the gradient is negative and higher than -79 $km^{-1}$, in red when  super-refraction  allowed reception. After the gradient reached  -115 $km^{-1}$ (at an elevation of 0.8 km) the wave bent downward and did not reach higher elevations.}
  \Description{Refractivity vs height at Cuneo station.}
  \label{cuneoRefra}
\end{figure}

Cuneo is another gateway reached from  the Gattinara probe, and its LOS  is also blocked by a sharp ridge, but nevertheless  packets  were received on February 2, 2020. Repeating the calculation for the geometry of this link, shown in Figure \ref{gattinara}, we get  $\nu$ =  4.81. Entering this value in equation 2 we see an additional attenuation of 26 dB. Since the FSL  over 140 km is 134 dB, the total attenuation is 160 dB, implying a received signal of - 146 dB, below the receiver sensitivity. So, to explain the reception we used the data gathered by the radiosonde at Cuneo, 8 km away, to draw Figure \ref{cuneoRefra}. The red line corresponds to a day with packet reception, the green line corresponds to another day where no packets were received.
From the red line we can see a condition of super-refraction, with  $\Delta$N/$\Delta$h = -115 at 0.8 km. This corresponds to K = 3.74, and entering this value in BotRf we obtain the profile of Figure \ref{gattinara}, in which the geometric parameter \textit{$\nu$} can be calculated as 2.07. Entering this in equation 2 produces a difraction attenuation of 19 dB, so now the received level is 14-(134+19) = -139 dBm, which is indeed detectable. We conclude that there is a \textbf{combination of difraction and super-refraction} in this case.




\section{Future work}

We intend to add more TPs in the near future. We would like to engage the TTN community to add more devices to the platform. Devices that are used for other purposes can be monitored by the TROPPO platform. In order to do so, we need access to the metadata of the data packets and not to the data payload itself. Please contact the corresponding author of this paper if you are interested in collaborating in such experiment.

\section{Conclusions}
We have presented a novel platform to obtain a great number of instances of anomalous propagation by leveraging the existing TTN gateways infrastructure. Deploying a few probes that periodically transmit short frames allows the detection of tropospheric propagation phenomena and also sheds light in the specific kind of anomalous propagation involved in each case, by leveraging  meteorological publicly available data from radiosondes. We have developed a tool that automatically sorts out paths in which the line of sight is blocked, so that the propagation can only be accounted for by diffraction,  super-refraction, tropospheric ducting or a combination of diffraction and super-refraction. Examples  are provided of each of the four kinds, confirmed by meteorological data gathered at nearby stations. The TP hardware deployed is inexpensive and by leveraging the  existing infrastructure of  a great number of TTN gateways and public meteorological data, it can be easily scaled up  by simply adding additional end-nodes.
\begin{acks}
We would like to thank; Sebastian Buettrich and John Cassidy for the constructive discussions; CISAR Trieste for the kind support;  TTN for the fantastic  community they have created. 

Data from the metereological station were obtained from \cite{uwyo} of the University of Wyoming, curated by Larry Oolman to whom we express our gratitude.

\end{acks}

\bibliographystyle{ACM-Reference-Format}
\bibliography{sample-base}

\end{document}